\documentclass[aps,pre,twocolumn,groupedaddress]{revtex4-1}

\usepackage{graphicx}% Include figure files
\usepackage{dcolumn}% Align table columns on decimal point

\usepackage{bm}% bold math
\usepackage{amsfonts}
\usepackage{amsmath}

\usepackage[dvips]{color}

\def\be{\begin{eqnarray}}
\def\ee{\end{eqnarray}}

\def\ba{\begin{array}}
\def\ea{\end{array}}

\def\mrm{\mathrm}

\def\del{\partial}
\def\vep{\varepsilon}

\def\pb{\overline{p}}
\def\qb{\overline{p}\,}

\def\p1{\dot{p}}

\def\qsn{p_\mrm{sn}}
\def\psn{p^*_\mrm{sn}}

\def\psisn{\psi_\mrm{c}}

\def\dTn{T'_n(1)}
\def\dVn{\overline{T}'_n(1)}

\def\Mn{\mathrm{M}_n}
\def\cn{\vec{c}_n}

\definecolor{darkgreen}{rgb}{0.0, 0.5, 0.0}
\definecolor{purple}{rgb}{0.50, 0.0, 0.50}

\begin{document}

% \preprint{}

\title{
Transition-type change between an inverted Berezinskii-Kosterlitz-Thouless transition and an abrupt transition 
in the bond percolation on a random hierarchical small-world network
}

\author{Tomoaki Nogawa}
\email{nogawa@med.toho-u.ac.jp}
\affiliation{%
Faculty of Medicine, Toho University, 
5-21-16, Omori-Nishi, Ota-ku, Tokyo 143-8540, Japan
}

\author{Takehisa Hasegawa}
\email{hasegawa@m.tohoku.ac.jp}
\affiliation{%
Graduate School of Information Science, 
Tohoku University, 
6-3-09, Aramaki-Aza-Aoba, Sendai, Miyagi 980-8579, Japan
}

\date{\today}

% phase transition, percolation, renormalization group, networks

\begin{abstract}
We study the bond percolation on a one-parameter family of a hierarchical small-world network, 
and find the metatransition between an inverted Berezinskii-Kosterlitz-Thouless (iBKT) transition 
and an abrupt transition driven by changing the network topology. 
It is found that the order parameter is continuous and the fractal exponent is discontinuous in the iBKT transition, 
and oppositely, the former is discontinuous and the latter is continuous in the abrupt transition. 
The gaps of the order parameter and the fractal exponent in each transition vanish as they approach the metatransition point. 
This point corresponds to a marginal power-law transition. 
In the renormalization group formalism, this metatransition corresponds to the transition between 
transcritical and saddle-node bifurcations of the fixed point via a pitchfork bifurcation. 
\end{abstract}

\maketitle
%%%%%%%%%%%%%%%%%%%%%%%%%%%%%%%%%%%%%%%%%%%%%%%%%%%%%%%%%%%%%%%%%%%%%%%%%%%%%%%
\section{Introduction}

Recently, cooperative phenomena (e.g. percolation and ferromagnetism) 
on various kind of non-Euclidean graphs 
have been investigated extensively in the context of complex networks \cite{Albert2002}, 
and a lot of exotic behaviors have been found \cite{Dorogovtsev08}. 
One of the most remarkable facts is that such a system often exhibits a critical phase 
\cite{Nogawa-Hasegawa09}, 
inside which a system always shows properties typical 
for ordinary critical {\it points} of the second-order transition 
such as infinite susceptibility and a zero-order parameter \cite{Hinczewski06}. 
This phase lies between an ordered phase 
(e.g., a percolating phase and a ferromagnetic phase) 
and a disordered phase (e.g., a nonpercolating phase and a paramagnetic phase); 
it is not a boundary but it occupies finite fraction of the parameter space. 
Although such a phase is similar to the quasi-long-range-order phase 
of the two-dimensional $XY$ model, \cite{Berezinskii72, Kosterlitz73, Kosterlitz74}, 
the origin is quite different. 
Some studies imply that a critical phase is attributed to 
a small-world property of graphs \cite{Hinczewski06, Nogawa-Hasegawa09b, Hasegawa-Sato2010}.

Recognition of a critical phase leads us to have interest in a transition 
between critical and noncritical phases. 
To our knowledge, disorder-critical transition has been found 
only in nonamenable graphs such as trees, 
and the property of its singularity has been well investigated 
\cite{Muller-Hartmann74, Schonmann2001, Nogawa-Hasegawa09}.
On the other hand, critical-order transition has been found in various
systems, and its singularity is distinguished by the behavior of the
order parameter near the phase boundary as the following: 
(T1) power-law transitions \cite{Hasegawa-Sato2010}, 
(T2) inverted Berezinskii-Kosterlitz-Thouless (iBKT) transitions 
\cite{Krapivsky04, Bauer05, Bollobas05, Riordan05, Hinczewski06} and 
(T3) abrupt transitions \cite{Nogawa-Hasegawa09, Boettcher12}. 
The order parameter $m$ varies with the distance from the phase boundary $\vep$ in the ordered phase as 
$m \propto \vep^\beta$ in T1, 
$m \propto \exp[-\alpha/\vep^{1/2}]$ in T2, and 
$m = m_c + c_1 \vep^b$ in T3, respectively.
Here $\beta$, $\alpha$, $m_c$, $c_1$ and $b$ are constants.

On the other hand, singularity cannot be quantified 
by the $\vep$ dependence of $m$ or the susceptibility $\chi$ 
in a critical phase, where $m=0$ and $\chi=\infty$ in an infinite size system. 
The authors proposed the characterization of the singularity by the fractal exponent $\psi$ 
\cite{Nogawa-Hasegawa09b}. 
It is related to the finite size dependence of a local disconnected susceptibility \cite{Bauer05}
of a specific site $\tilde{\chi}$, 
which includes the contribution of the long-range coherence 
and corresponds to the size of the cluster that contains the focused site in percolation models.
In highly inhomogeneous graphs such as nonamenable graphs, 
where the boundary effect is not negligible even in the thermodynamic limit, 
and scale free networks, 
the local disconnected susceptibility of a central site or hub can diverge 
even when the spatially averaged susceptibility does not. 
Therefore the critical phase of such a system is characterized better by the former 
\cite{Nogawa-Hasegawa09, Hasegawa10c, Hasegawa-Sato2010, Boettcher12, Nogawa-Hasegawa12, Nogawa-Hasegawa12b}.
We introduce $\psi$ such that $\tilde{\chi} \propto N^\psi$, where $N$ is the number of degrees of freedom. 
Whereas $\psi$ takes trivial values: 0 in a disordered phase and 1 in an ordered phase, 
it takes a fractional value between 0 and 1 in a critical phase. 
Previous studies on several systems indicate that the continuity of $\psi$ 
at critical-order transitions is opposite to that of $m$; 
$\psi$ continuously approaches 1 as $1 - \psi \propto \vep^{\nu}$ in T3 \cite{Nogawa-Hasegawa09, Boettcher12} 
whereas it continuously approaches an edge value $\psi_c < 1$ from below 
and jumps to 1 in T1 and T2 \cite{Hasegawa-Sato2010}.

At this moment, we lack a unified picture of various critical-order transitions. 
Recently, all of the aforementioned transitions were found in different models on the same graph, a hierarchical small-world network (HSWN) with a one-dimensional backbone, 
known as the Farey graph \cite{Zhang-Comellas2011}: 
T2 for the $Q$-state Potts model with $Q \ge 3$ \cite{Nogawa-Hasegawa12}, 
T1 for the Potts model with $Q=2$ \cite{Nogawa-Hasegawa12b} 
and T3 for bond percolation \cite{Boettcher12}. 
Note that bond-percolation can be mapped to the Potts model with $Q=1$. 
The authors predicted a metatransition by changing $Q$ at $Q_c=2$ \cite{Nogawa-Hasegawa12b}. 
Here metatransition means the transition between transitions T2 and T3 via T1 as a metatransition point. 
Although this metatransition can be a clue 
to establishing a comprehensive theory of the critical-order transitions, its property is rarely understood 
due to the fact that the number of spin state $Q$, which is an integer, cannot be changed continuously. 
In this paper, we investigate the metatransition in detail 
by analyzing the bond percolation on a random HSWN whose topology can be changed by a {\it continuous} parameter.

\section{Structure of the graph}

%%%%%%%%%%%%%%%%%%%%%%%%%%%%%%%%%%%%%%%%%%%%%%%%%%%%%%%%%%%%%%%%%
\begin{figure}[t]
% \hspace{0.1cm}{\bf (a)}\hspace{7.05cm}{\bf (b)}\\ \vspace{-1.2cm}
\begin{center}
% \hspace{-6.8cm} {\bf{\large (a)}}\\ \vspace{-14.1pt}
\includegraphics[trim=60 590 130 -220,scale=0.4,clip]{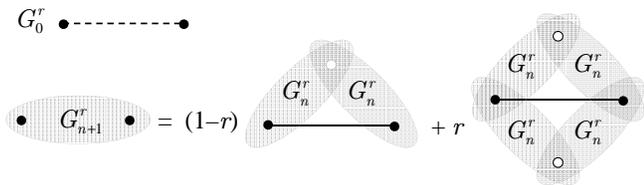}
\end{center}
\vspace{-5mm}
\caption{\label{fig:graph}
Initial graph $G_0^r$ and schematic diagram of the recursive construction of the hierarchical graph. 
The solid circles are the roots of both $G_n^r$ and $G_{n+1}^r$, 
and the empty circles are the roots of $G_n^r$ and not of $G_{n+1}^r$. 
The broken and solid lines represent the backbone and shortcut edges, respectively.
}
\end{figure}
%%%%%%%%%%%%%%%%%%%%%%%%%%%%%%%%%%%%%%%%%%%%%%%%%%%%%%%%%%%%%%%%%

% \vspace{1mm} \noindent {\it Structure of Graph }:
Here we explain how to construct the sequence of random graphs that we treat. 
We start with $G^r_0$, which consists of two vertices, 
which are called ``roots,'' connected by a ``backbone edge.'' 
As illustrated in Fig.~\ref{fig:graph}, 
we recursively make the graph $G^r_{n+1}$ from two $G^r_n$'s with probability $1-r$, 
and from four $G^r_n$'s with probability $r$. 
Here $G^r_n$'s are independent random realizations. 
In both cases, we make graph operations: joining pairs of root vertices in $G^r_n$'s to be one, 
letting two of the roots in $G^r_{n}$'s be the roots of $G^r_{n+1}$, 
and adding a ``shortcut edge'' connecting the new roots. 
The deterministic cases with $r=0$ and $r=1$ coincide with the Farey graph 
\cite{Boettcher12,Nogawa-Hasegawa12,Nogawa-Hasegawa12b} 
and the decorated (2,2)-flower \cite{Hinczewski06, Rozenfeld07, Berker09, Hasegawa-Sato2010}, respectively. 
The expectation value of the number of vertices in $G_n^r$ increases as 
\be
\langle N_{n+1} \rangle = \big\langle k \langle N_n \rangle - l \big\rangle, 
\ee
where $\langle \cdots \rangle$ means the average over graph realization. 
% and we used the independence among $G^r_n$'s. 
%  up to $n$th generation 
% while $\langle \cdots \rangle$ means the full average up to $(n+1)$th generation. 
Random variables $(k, l)$ equal (2,1) with probability $1-r$ and (4,4) with probability $r$. 
Then we have $\langle N_n \rangle = [ ( 1 + r ) \kappa^n   + ( 1 + 3r) ]/(1+2r)$, 
and $\langle N_n \rangle$ is approximately proportional to $\kappa(r)^n$ for $n \gg 1$, 
where $\kappa(r)  \equiv  2(1+r)$. 
We call the subgraph obtained by removing all shortcut edges the backbone, 
where the shortest path length between the two roots equals $L_n = 2^n$.
The effective dimension of the backbone $d_\mrm{eff}$ such that 
$\langle N_n \rangle \propto (L_n)^{d_\mrm{eff}}$
 is given by $d_\mrm{eff} = \log_2 \kappa(r)$, 
which monotonically changes $1 \to 2$ with $r:0 \to 1$. 
With the shortcut edges, the graph is small-world, i.e. infinite-dimensional, 
in the sense that the shortest path length is proportional to $\ln N_n$.

\section{Renormalization of the percolating probability}

% \vspace{1mm} \noindent {\it Renormalization of percolating probability}: 
We consider the bond percolation on $G^r_n$. 
Let $p$ be the open-bond probability for both the backbone and shortcut edges. 
By utilizing the hierarchical structure of $G^r_n$, 
we can derive an exact renormalization group (RG) map of $p$.
We introduce the probability $p_n$ that the two roots of $G^r_n$ are connected; 
i.e., they belong to the same cluster. 
Its average over graph realizations satisfies the recursion equation starting from $p_0 = p$: 
\begin{eqnarray}
&& p_{n+1} 
% &=& \rb [ 1 - (1-p) (1-p_n^2)  ] + r [ 1 - (1-p) (1-p_n^2)^2] 
= 1 - (1-p)(1-p_n^2)(1-r p_n^2), 
\nonumber \\
&& \Leftrightarrow p_{n+1} - p_n = ( 1 - p_n) [ p( 1 + A(r, p_n) ) - A(r, p_n) ], 
\label{eq:rg_diff}
% \\
% &&\mrm{where} \quad 
% \label{eq:rg_pn}
\\
&& A(r, p) \equiv p [ 1 - r p (1 + p) ]. 
\end{eqnarray} 
Hereafter, $p_n$ denotes the averaged value over graph realizations. 
The nontrivial fixed point (NTFP) for given $r$ and $p$, 
$p^*(r, p) \in (0, 1)$, is obtained by solving 
\begin{eqnarray}
% q = \frac{A(r, p^*)}{1+A(r, p^*)} \iff 
A(r, p^*) = p/(1-p). 
\label{eq:A-q}
\end{eqnarray}

%%%%%%%%%%%%%%%%%%%%%%%%%%%%%%%%%%%%%%%%%%%%%%%%%%%%%%%%%%%%%%%%%
\begin{figure}[t]
% \hspace{0.1cm}{\bf (a)}\hspace{7.05cm}{\bf (b)}\\ \vspace{-1.2cm}
\begin{center}
% \hspace{-6.8cm} {\bf{\large (a)}}\\ \vspace{-14.1pt}
\includegraphics[trim=40 20 220 20,scale=0.22,clip]{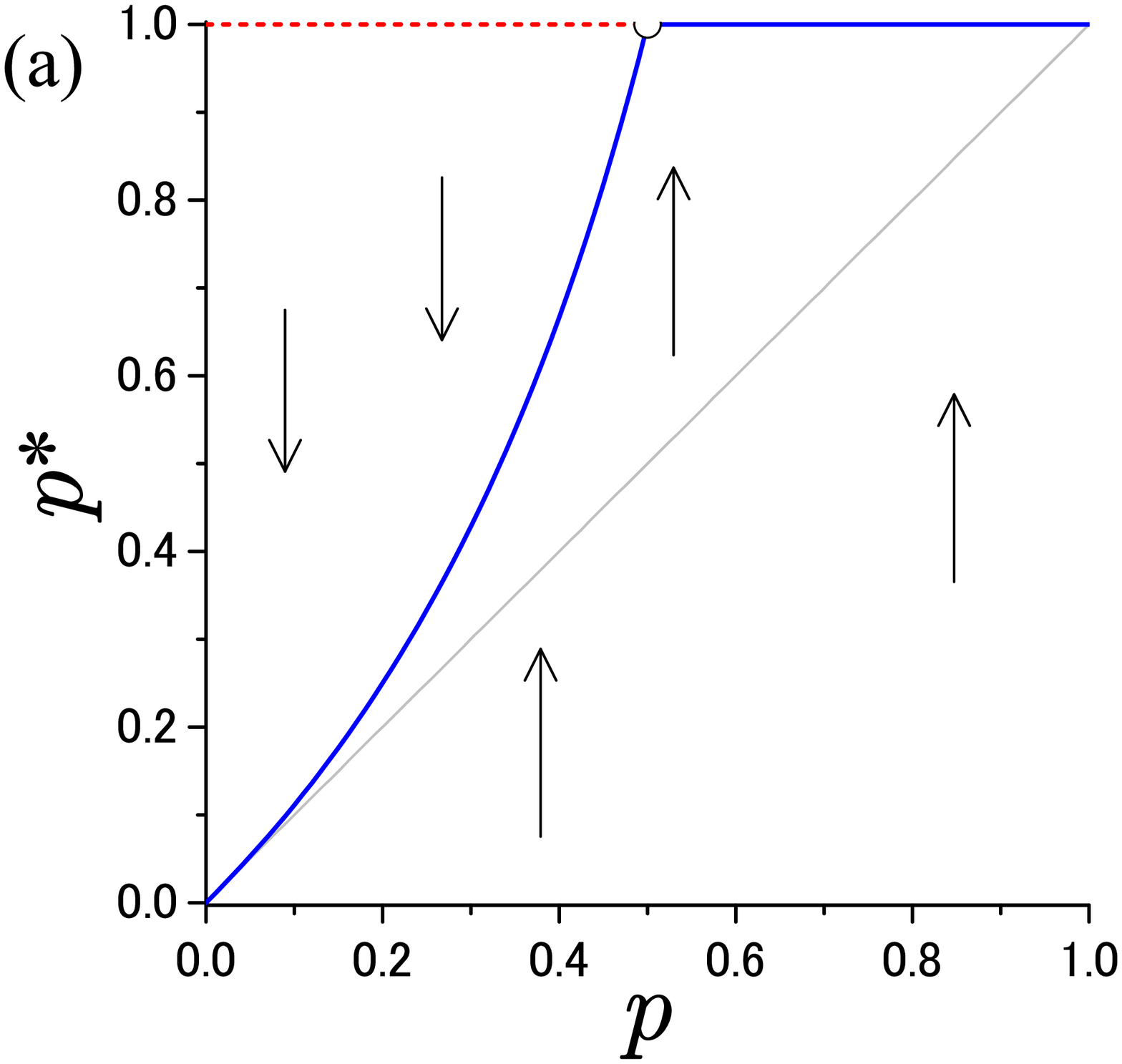}
\hspace{-2mm}
\includegraphics[trim=40 20 220 20,scale=0.22,clip]{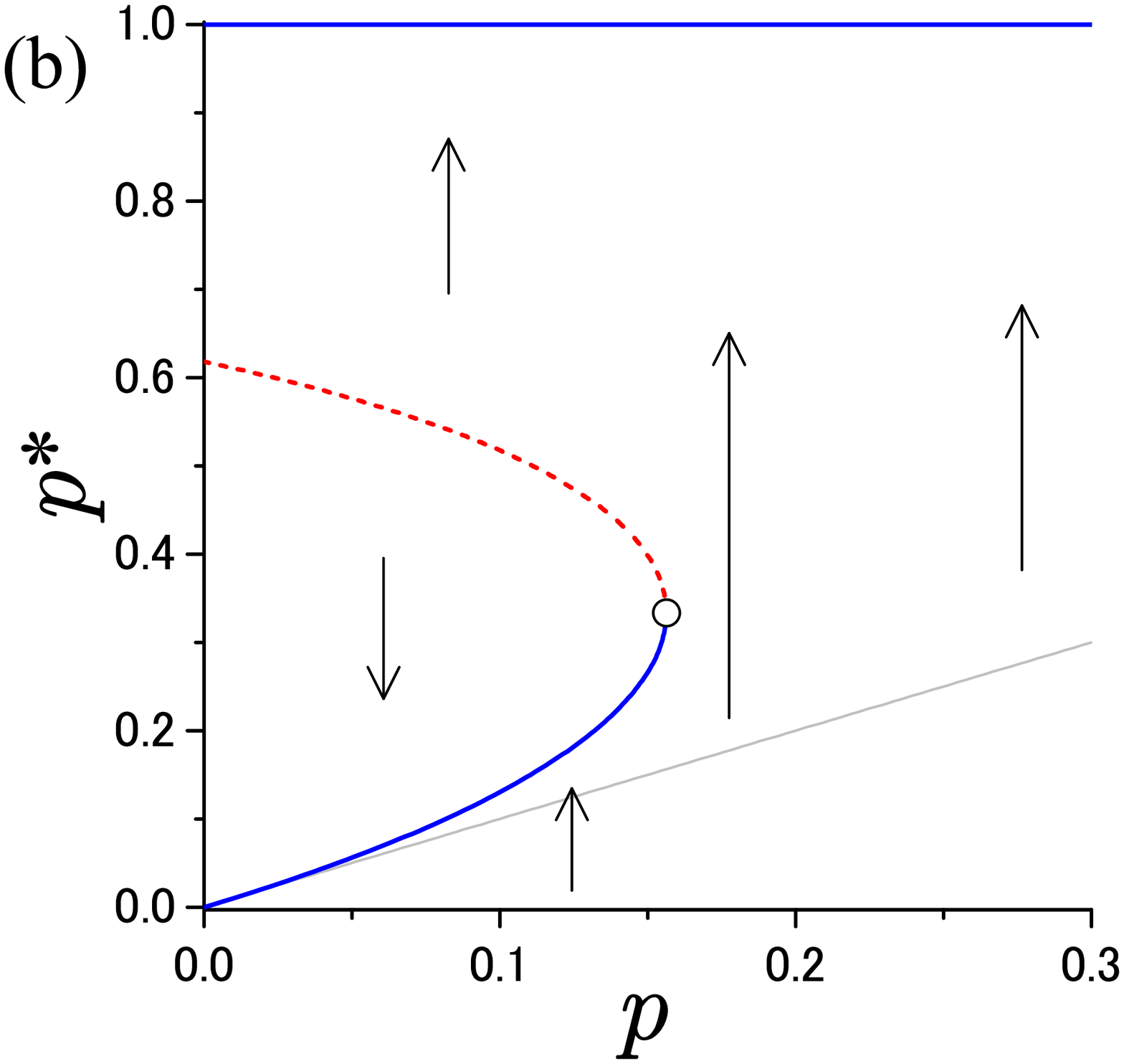}
\end{center}
\vspace{-5mm}
\caption{\label{fig:PD01}
(Color online) 
Renormalization group flow diagram for (a) $r=0$ and (b) $r=1$. 
The solid (blue) and broken (red) lines represent stable and unstable fixed points, respectively. 
The thin gray lines represent $p^* = p$, which is the starting point of the RG flow. 
}
\end{figure}
%%%%%%%%%%%%%%%%%%%%%%%%%%%%%%%%%%%%%%%%%%%%%%%%%%%%%%%%%%%%%%%%%

% \subsection{the case of $r=0$ (HSWN)}

Figure~\ref{fig:PD01}(a) shows the RG flow for the case of $r=0$ (Farey graph). 
We consider the flow starts at $p_0=p$. 
As $p$ increases, a transcritical bifurcation of the stable fixed point occurs at $p=p_c = 1/2$; 
$p_n$ converges to the NTFP for $p < p_c$ and to 1 for $p \ge p_c$. 
Thus the system is in a critical phase for $p < p_c$ and in a percolating phase for $p > p_c$. 
A nonpercolating phase corresponding to $p^*=0$ does not exist. 
The above bifurcation corresponds to T2 as shown in Ref.~\cite{Boettcher12}.

% \subsection{the case of $r=1$ (decorated (2,2)-flower)}

For the case of $r=1$ [decorated (2,2)-flower], 
there exist two NTFPs for $p < \psn = 5/32$ as shown in Fig.~\ref{fig:PD01}(b).
The one with smaller $p^*$ is stable and the other is unstable. 
The two NTFPs meet and annihilate together at the saddle-node bifurcation point (SNBP), 
$(p,p^*) = (\qsn, p^*_\mrm{sn}) = (5/32, 1/3)$. 
Consequently, a saddle-node bifurcation occurs with increasing $p$; 
$p_n$ converges to the stable NTFP for $p<p_c=\qsn$ and to 1 for $p>p_c$. 
Therefore $p^*$ jumps from $p^*_\mrm{sn}$ to $p^*=1$ at $p_c$. 
This bifurcation corresponds to T3 as shown in Ref.~\cite{Berker09, Hasegawa-Sato2010}. 
[Note that the stable fixed point $p^*=1$ and the unstable fixed point for $p<p_c$ are irrelevant in the present case with $p_0 = p < p^*(r,p)$. ]

%%%%%%%%%%%%%%%%%%%%%%%%%%%%%%%%%%%%%%%%%%%%%%%%%%%%%%%%%%%%%%%%%
\begin{figure}[t]
% \hspace{0.1cm}{\bf (a)}\hspace{7.05cm}{\bf (b)}\\ \vspace{-1.2cm}
\begin{center}
% \hspace{-6.8cm} {\bf{\large (a)}}\\ \vspace{-14.1pt}
\includegraphics[trim=40 20 220 20,scale=0.22,clip]{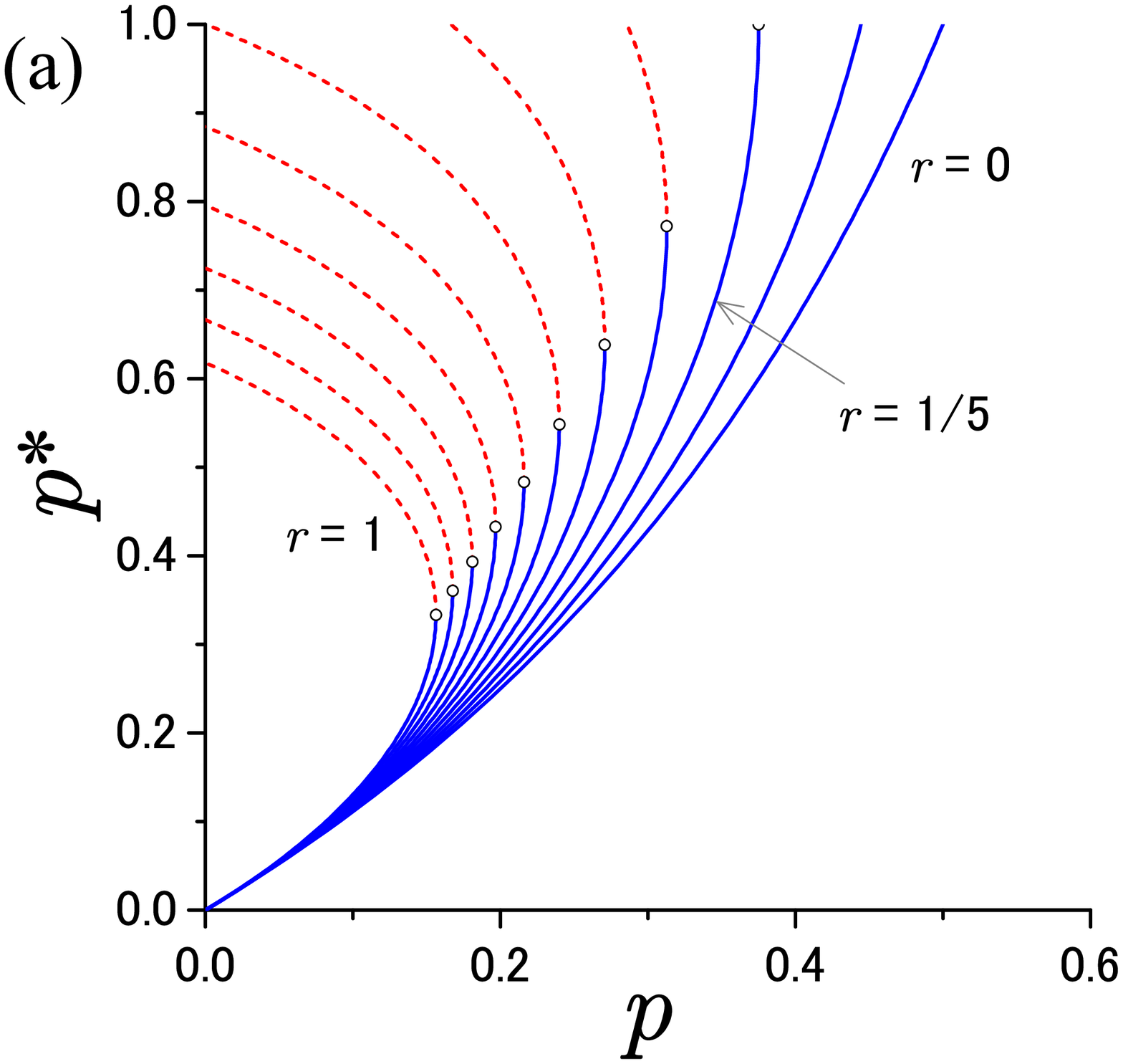}
\hspace{-2mm} 
\includegraphics[trim=40 20 220 20,scale=0.22,clip]{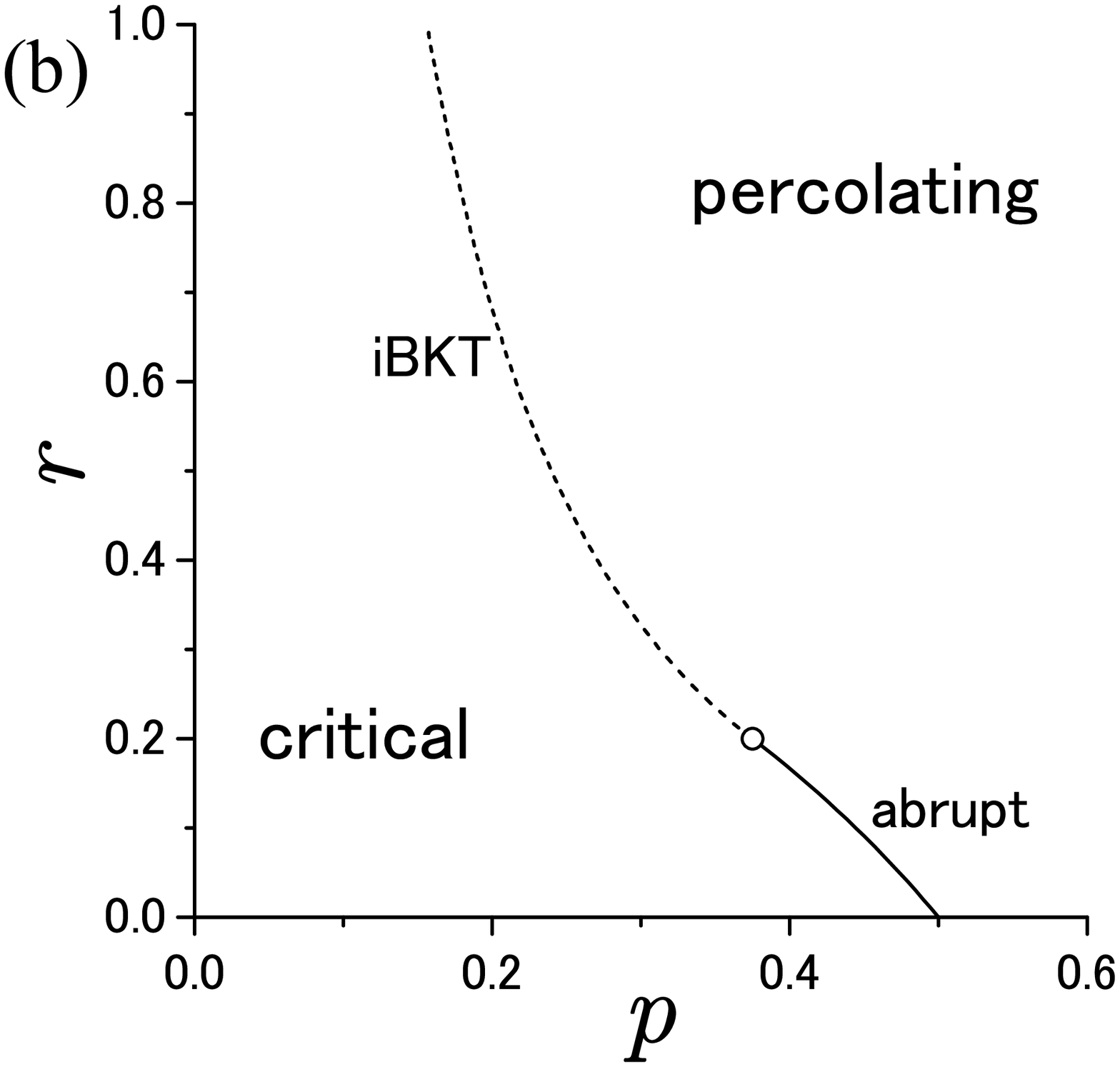}
\end{center}
\vspace{-5mm}
\caption{\label{fig:RG_p-q}
(Color online) 
(a) $p$ dependence of the nontrivial fixed points for $r=0.0-1.0$ (0.1 step).
The solid (blue) and broken (red) lines represent stable and unstable fixed points, respectively.
(b) Phase diagram in the $r \times p$ plane. 
The abrupt transition occurs across the solid line 
and the iBKT transition does across the broken line.
}
\end{figure}
%%%%%%%%%%%%%%%%%%%%%%%%%%%%%%%%%%%%%%%%%%%%%%%%%%%%%%%%%%%%%%%%%

% \subsection{the case of general $r$}

Next we consider the case of general $r \in [0, 1]$. 
Figure~\ref{fig:RG_p-q}(a) shows the $p$ dependence of the NTFP for several values of $r$. 
The type of bifurcation of the RG fixed point changes at $r=r_c=1/5$; 
saddle-node type for $r > r_c$ and transcritical type for $r \le r_c$. 
The SNBP is given by $\del p / \del p^* = 0$, which leads to 
\begin{eqnarray}
\big( \qsn(r), \psn(r) \big)
= \left(
\frac{A(r, \psn(r)) }{1 + A(r, \psn(r)) }, \ 
\frac{\sqrt{1+3/r} - 1 }{3} 
\right).
\end{eqnarray}
As $r$ decreases, $\psn$ becomes larger to be unity at $r=r_c$, 
and the SNBP enters the unphysical region $p^* > 1$. 
At $r=r_c$, a pitchfork bifurcation occurs by changing $p$. 
Equation~\eqref{eq:A-q} leads to the $p$ dependence of the {\it stable} fixed points 
near and below $p_c$ for given $r$ as 
\begin{eqnarray}
1 - p^*(r,p) =
\left \{
\ba{ccc}
|\vep| & \mrm{for} & r < r_c, 
\\
\sqrt{|\vep|} & \mrm{for} & r = r_c, 
\\
\, [1 - \psn(r) ] + \sqrt{|\vep|} & \mrm{for} & r > r_c.
\ea
\right.
\label{eq:law_p-q}
\end{eqnarray}
Here $\vep \equiv p - p_c(r)$ and we omit unimportant coefficients of the power of $|\vep|$ 
[the same as in Eqs.~\eqref{eq:law_psi-q} and \eqref{eq:law_P0-q} appearing later].
In the limit $r \to r_c + 0$, the gap of $p^*$ at $p_c$, i.e., $1 - \psn$, continuously approaches 0 as  
\be
1 - \psn(r) = (25/8) ( r - r_c ) + O( (r-r_c)^2 ).
\label{eq:law_p-r}
\ee

Figure.~\ref{fig:RG_p-q}(b) shows the phase diagram in the $r \times p$ plane. 
The threshold probability $p_c(r)$ is given by $\qsn(r)$ for $r > r_c$ 
and by $[ 1 + A^{-1}(r,1)]^{-1}$ for $r < r_c$.  
As shown next, the point $(p,r)=(p_c(r_c), r_c)$ plays a similar role with a tricritical point \cite{Griffiths1973}, 
which joins a first-order transition boundary and a second-order one.

\section{Generating function analysis}

% \vspace{1mm} \noindent {\it Generating function analysis}: 
% \subsection{recursion equations}
We investigate the order parameter and the fractal exponent of the root cluster, 
i.e., the cluster that one root belongs to. 
The size of the root cluster of a given random graph averaged over the open-bond realization $s^\mrm{rt}_n$ 
can be calculated by using the generating function 
$Z_n(x) = \sum_s z_{n,s} x^s$, 
where $z_{n,s}$ is the probability that a root belongs to a cluster with size $s$. 
We split it in two terms as $Z_n(x) = x^2 T_n(x) + x \overline{T}_n(x)$ with 
$T_n(x) = \sum_{s=0}^{\infty} t_{n,s} x^s$
and 
$\overline{T}_n(x) = \sum_{s=0}^{\infty} \overline{t}_{n,s} x^s$, 
where $t_{n,s}$ is the probability that 
a root belongs to a cluster with size $s+2$ that includes another root, 
and $\overline{t}_{n,s}$ is the probability that 
a root belongs to a cluster with size $s+1$ that does not include another root \cite{Hasegawa-Sato2010}. 
The expectation value of a root cluster size is given by 
\begin{equation}
s_n^\mrm{rt}  =  Z'_n(1) 
= [ 2 T_n(1) + \overline{T}_n(1)] + [ \dTn + \dVn ], 
\label{eq:srt}
\end{equation}
where the primes indicate the derivatives with respect to $x$, 
and $T_n(1) = 1 - \overline{T}_n(1)  = p_n$. 
The recursion equations of $\dTn$ and $\dVn$ 
averaged over realizations of random graphs are obtained 
by utilizing those for $r=0$ in Ref.~\cite{Boettcher12} and for $r=1$ in Ref.~\cite{Hasegawa-Sato2010} as 
\be
\langle \vec{\tau}_{n+1} \rangle
= \Big\langle
\Mn \langle \vec{\tau}_n \rangle + \cn
\Big\rangle, 
\quad 
\vec{\tau}_n \equiv 
\left( \ba{c}
\dTn \\ \dVn
\ea \right) .
\label{eq:tau}
\ee
Here $( \Mn, \cn )$ 
equals $( \mrm{M}_n^{(0)}, \vec{c}_n^{\,(0)} )$ with probability $1-r$ 
and $(\mrm{M}_n^{(1)}, \vec{c}_n^{\,(1)})$ with probability $r$ where 
\be
&
\mrm{M}_n^{(0)} 
=
\left( \ba{cc}
2[ \qb p_n + p ] & 2 p \p1_n
\\
\qb  \pb_n & \qb \p1_n
\ea \right) , \ 
\vec{c}_n^{\,(0)} 
=
\left( \ba{c} 
p_n [ p_n + 2 p \pb_n ]
\\
\qb p_n \pb_n
\ea \right) , 
\nonumber \\ &
\mrm{M}_n^{(1)}
=
\left( \ba{cc}
4[ 1  - \qb  \pb_n^2 \p1_n]  &  4 \p1_n [ p_n^2 + p \pb_n  \p1_n ]
\\
2 \qb  \pb_n^2 \p1_n  &  2 \qb  \pb_n \p1_n^2
\ea \right) , 
\nonumber \\ &
\vec{c}_n^{\,(1)} 
= 
\left( \ba{c} 
2 p_n [ 2 - p_n - 2 \qb  \pb_n^2 \p1_n ]
\\
2 \qb p_n \pb_n^2 \p1_n
\ea \right) , 
\\
&
% \rb \equiv 1-r, \ 
\qb \equiv 1-p, \ 
\pb_n \equiv 1 - p_n, \ 
\p1_n \equiv 1 + p_n.
\end{eqnarray}

%%%%%%%%%%%%%%%%%%%%%%%%%%%%%%%%%%%%%%%%%%%%%%%%%%%%%%%%%%%%%%%%%
\begin{figure}[t]
% \hspace{0.1cm}{\bf (a)}\hspace{7.05cm}{\bf (b)}\\ \vspace{-1.2cm}
\begin{center}
\includegraphics[trim=60 35 40 30,scale=0.3,clip]{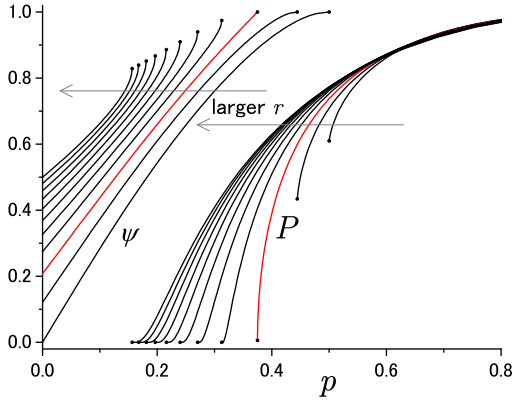}
\end{center}
\vspace{-5mm}
\caption{\label{fig:X-q}
(Color online) 
$p$ dependence of fractal exponent $\psi$ for $p \le p_c(r)$ 
and order parameter $P$ for $p \ge p_c(r)$ for $r=0.0-1.0$ (0.1 step). 
The red (grey) lines indicate the data at $r = r_c$.
We show the numerical data with $n=65536$ for $P$, 
in which the finite size effect is negligible in this plot.
}
\end{figure}
%%%%%%%%%%%%%%%%%%%%%%%%%%%%%%%%%%%%%%%%%%%%%%%%%%%%%%%%%%%%%%%%%

\subsection{Fractal exponent in the critical phase}

First, we consider the singularities of the fractal exponent. 
We have 
\be
\langle s_n^\mrm{rt} \rangle \propto [\lambda(r,p^*)]^n
\ee
for large $n$, 
where $\lambda(r,p)$ is the larger eigenvalue of 
$\mrm{M}_n^{(r)} = (1-r)\mrm{M}_n^{(0)} + r \mrm{M}_n^{\,(1)}$ 
with $p_n = p^*(r,p)$.
Then the fractal exponent $\psi(r,p)$ such that 
\be
\langle s_n^\mrm{rt} \rangle \propto \langle N_n \rangle ^{\psi(r,p)} 
\ee
is given by 
\be
\psi(r,p) = \ln \lambda(r,p) / \ln \kappa(r). 
\ee
The $p$ dependence of $\psi$ for several values of $r$ is shown in Fig.~\ref{fig:X-q}. 
For given $r$, $\psi$ increases with $p$ for $p < p_c(r)$ and equals 1 for $p > p_c(r)$. 
Near and below $p=p_c(r)$, we numerically found that $1-\psi$ obeys the power-law 
whose exponent depends on $r$ as 
\begin{eqnarray}
1 - \psi(r, p) =  
\left \{
\ba{ccc}
|\vep|^2 & \mrm{for} & r < r_c, 
\\
|\vep| & \mrm{for} & r = r_c, 
\\
\, [1 - \psisn(r) ] + \sqrt{|\vep|} & \mrm{for} & r > r_c, 
\ea
\right.
\label{eq:law_psi-q}
\end{eqnarray}
where $\psi_c(r) \equiv \psi(r, p_c(r))$.

\subsection{Order parameter in the percolating phase}

Next, we consider the singularities of the order parameter 
\be
P \equiv \langle s_n^\mrm{rt}  / N_n \rangle, 
\ee 
which means the probability that a randomly chosen vertex belongs to the root cluster.
To calculate $P$, we solve the following recursion equations: 
\be
\left \langle \frac{\vec{\tau}_{n+1} }{N_{n+1}} \right \rangle
&=& \left \langle \frac{ 
\Mn (N_n^{-1} \vec{\tau}_n) + N_n^{-1} \cn 
}{
k - l N_n^{-1}
}\right \rangle
\nonumber \\
&\approx& \left\langle \frac{ 
\Mn \langle N_n^{-1} \vec{\tau}_n \rangle + \langle N_n^{-1} \, \rangle \cn 
}{
k  - l \langle N_n^{-1} \rangle 
}\right\rangle , 
\label{eq:P_rec1}
\\
\langle N_{n+1}^{-1} \rangle &=& 
\left \langle \frac{N_n^{-1}}{k - l N_n^{-1}} \right \rangle 
\approx \left \langle \frac{ \langle N_n^{-1} \rangle }
{k - l \langle N_n^{-1} \rangle } \right \rangle. 
\label{eq:P_rec2}
\ee
In these formulas, we make an approximation to take the average of $N_n^{-1}$ 
in the denominator in advance, which is good for large $N_n$. 
The $p$ dependence of $P$ for $p > p_c(r)$ in Fig.~\ref{fig:X-q} 
is obtained by solving Eq.~\eqref{eq:P_rec1} and \eqref{eq:P_rec2} numerically 
and using Eqs.~\eqref{eq:srt} and \eqref{eq:tau}. 
Whereas $P$ for $n \to \infty$ equals 0 for $p < p_c$, 
$P$ is finite and increases with $p$ above $p_c$. 
Near and above $p_c(r)$, we numerically found three types of singular behavior for $P$ depending on $r$ as 
\begin{eqnarray}
P(r,p) = 
\left \{
\ba{ccc}
P_c(r) + \vep & \mrm{for} & r < r_c
\\
\vep^{\beta} & \mrm{for} & r = r_c
\\
\exp[ - \alpha(r)/\sqrt{\vep}\, ] & \mrm{for} & r > r_c
\ea
\right.
, 
\label{eq:law_P0-q}
\end{eqnarray}
where $P_c(r) = P(r,p_c(r))$. 
We estimate $\beta = 1/2$, which coincides with $\beta$ for the Potts model on the Farey graph 
with $Q=Q_c=2$ \cite{Nogawa-Hasegawa12b}. 
% But this coincidence may indicates the lack of universality when we recall $P=s/N$ corresponds to $\chi/N=m^2$. 

%%%%%%%%%%%%%%%%%%%%%%%%%%%%%%%%%%%%%%%%%%%%%%%%%%%%%%%%%%%%%%%%%
\begin{figure}[t]
% \hspace{0.1cm}{\bf (a)}\hspace{7.05cm}{\bf (b)}\\ \vspace{-1.2cm}
\begin{center}
\includegraphics[trim=80 100 70 30,scale=0.26,clip]{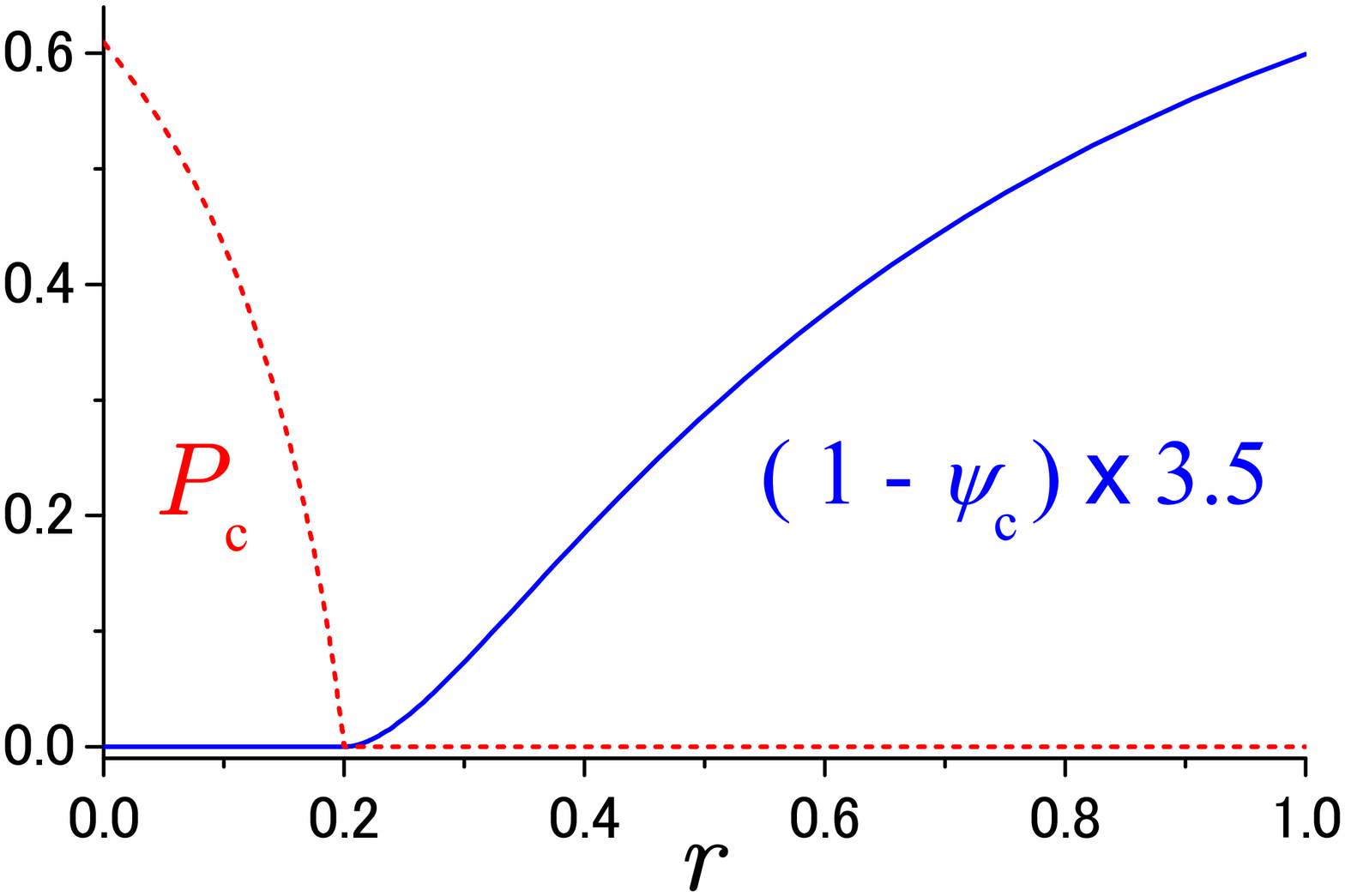}
\end{center}
\vspace{-5mm}
\caption{\label{fig:gap}
(Color online) $r$ dependence of $1 - \psi_c$ (blue solid line) 
and $P_c$ (red broken line).
}
\end{figure}
%%%%%%%%%%%%%%%%%%%%%%%%%%%%%%%%%%%%%%%%%%%%%%%%%%%%%%%%%%%%%%%%%

\subsection{Metaorder parameters} 

Figure~\ref{fig:gap} shows the $r$ dependence of the gaps $1-\psi_c$ and $P_c$. 
Whereas $1-\psi_c$ is zero for $r \le r_c$ and finite for $r>r_c$, 
$P_c$ is zero for $r \ge r_c$ and finite for $r<r_c$.
These gaps play the role of an order parameter in the metatransition. 
We actually found the following power-laws: 
\be
1 - \psisn(r) \propto ( r - r_c )^2 & \quad  \mrm{for} \ \  & r > r_c,   
\label{eq:law_psi-r}
\\
P_c(r) \propto  r_c - r  & \quad \mrm{for} \ \ & r < r_c. 
\label{eq:law_P0-r}
\ee
The power exponents 2 and 1 are numerically estimated with good precision.

\section{Conclusions}

In this paper, we investigated the metatransition 
in the bond percolation on a random hierarchical small-world network  
whose topology could be varied by the continuous parameter $r$. 
It is clearly shown that the metatransition at $r_c$ from the iBKT transition to the abrupt transition 
corresponds to the event where the SNBP gets over the ordered fixed point to enter the unphysical region. 
A similar classification is proposed by Boettcher and Brunson \cite{Boettcher11, Boettcher-Brunson11}. 
At the transition for $r<r_c$, which is governed by a transcritical bifurcation point, 
the order parameter is discontinuous and the fractal exponent is continuous. 
At the transition for $r>r_c$, which is governed by a saddle-node bifurcation point, 
the order parameter is continuous and the fractal exponent is discontinuous. 
At the marginal transition at $r=r_c$, which is governed by a pitchfork bifurcation point, 
both the fractal exponent and the order parameter are continuous. 
The metatransition is characterized by a ``metaorder parameter'': 
the gap of the order parameter for $r<r_c$ and the gap of the fractal exponent for $r>r_c$, 
both of which continuously vanish as $r \to r_c \pm 0$, respectively. 
Universality of the present criticality of the phase transitions, Eqs.~(\ref{eq:law_psi-q}, \ref{eq:law_P0-q}),  
and that of the metatransition, Eqs.~(\ref{eq:law_psi-r}, \ref{eq:law_P0-r}), is an open question.

The continuity of $\psi$ is noteworthy because it is related to the divergence of a correlation length $\xi$ 
as $1 - \psi \propto \xi^{-1}$ in nonamenable graphs such as enhanced trees \cite{Nogawa-Hasegawa09b}, 
which have a small-world property in the sense that the dimension is proportional to $\ln N$. 
If we extend the concept of correlation length in general small-world systems by $\xi \propto (1-\psi)^{-1}$, 
the difference between the critical-order transitions and conventional phase transitions stands out. 
The order parameter $m$ is continuous and $\xi^{-1}$ is discontinuous at the iBKT transition for $r > r_c$ 
and $m$ is discontinuous and $\xi^{-1}$ is continuous at the abrupt transition for $r < r_c$. 
On the other hand, both $m$ and $\xi^{-1}$ are discontinuous at an ordinary first order transition 
and both of them are continuous at an ordinary second-order transition. 
Furthermore, we consider that the power-law transition at the metatransition point 
is not a generic critical-order transition with power-law (T1) but a special one at the marginal point. 
This transition is governed by a stable fixed point \cite{Nogawa-Hasegawa12b}, and $\psi$ is continuous therein, 
whereas a power-law transition is ordinarily governed by an unstable fixed point and $\psi$ is discontinuous therein. 
Such a generic power-law transition is observed, e.g., by extending the present model; 
a transition occurs at the unstable NTFP shown in Fig.~\ref{fig:RG_p-q} 
if different open-bond probabilities are given for the backbone and shortcut edges 
to provide $p_0 > \qsn(r,p)$ (see Ref.~\cite{Hasegawa-Sato2010}).

%Universality of the critical behavior of the phase transition and the metatransition is an open question.  
%Since most of the exponents appeared in Eqs.~(\ref{eq:law_p-q}, \ref{eq:law_p-r}, 
%\ref{eq:law_psi-q}, \ref{eq:law_psi-r}, \ref{eq:law_P0-q}, \ref{eq:law_P0-r}) 
%take integer or one half, they are expected to be universal values for each. 
%The relation between the exponents for different quantities is also a question 
%[e.g., it seems that $1 - \psi \propto ( 1 - p^* )^2$ ]. 

\section*{ACKNOWLEDGMENTS}

The authors thank S. Boettcher and K. Nemoto for fruitful discussions. 
This work was supported by 
Grants-in-Aid for Young Scientists (B) (Grants No.~24740054 and No.~25800214) 
and by the JST, ERATO, Kawarabayashi Large Graph Project.

% \bibliography{d:/home/study/manuscripts/mybib.bib}
% \input{MetaTransition.bbl}

% \bibliography{S:/manuscripts/mybib.bib}


\end{document}